\documentclass[9pt,twocolumn,twoside]{opticajnl}
\journal{opticajournal} % use for journal or Optica Open submissions

\setboolean{shortarticle}{true}
\usepackage{lineno}
\usepackage{siunitx}
\usepackage{todonotes}

\title{High-resolution rapid-scanning Fourier-transform spectroscopy of ultracold atoms}

\author[1]{Friedemann Landmesser}
\author[1]{Nicolai Gölz}
\author[1]{Ulrich Bangert}
\author[1]{Tobias Sixt}
\author[1,*]{Lukas Bruder}

\affil[1]{Institute of Physics, University of Freiburg, Hermann-Herder-Stra{\ss}e 3, 79104 Freiburg, Germany}
\affil[*]{lukas.bruder@physik.uni-freiburg.de}

\begin{abstract}
Femtosecond interferometry combined with acousto-optical phase modulation is an effective approach to implement various types of coherent nonlinear and multidimensional spectroscopy schemes. 
The high sensitivity of this method has recently enabled the study of highly dilute gaseous and ultracold quantum systems for which the attainable spectral resolution is of particular interest.  
Here, we directly compare the performance and spectral resolution between two experimental implementations, that are step-wise and continuous rapid scanning of the underlying Fourier transform interferometers. 
We show the performance advantage of the rapid-scanning approach and demonstrate a spectral resolution of 250\,MHz in the spectroscopy of laser-cooled Li atoms. 
This is a 10-fold resolution improvement compared to previous experiments. 
\end{abstract}

\setboolean{displaycopyright}{false} % Do not include copyright or licensing information in submission.

\begin{document}

\maketitle

\paragraph{Introduction}
Ultracold trapped atoms, molecules and ions provide important test systems to explore fundamental quantum phenomena\,\cite{georgescu_quantum_2014}. 
Traditionally these systems are studied with frequency-domain spectroscopy methods that can readily access the high spectral resolution attainable at ultracold temperatures. 
This is in contrast to nonlinear spectroscopy in the time domain incorporating femtosecond (fs) laser pulses that feature broad spectral bandwidth and thus low spectral resolution. 
However, the high intensity and short durations of fs laser pulses offer the unique possibility to design experiments capable of correlating and isolating very weak spectroscopic signatures. 
Popular examples are photon echo, multiple-quantum coherence (MQC) and coherent multidimensional spectroscopy (CMDS)\,\cite{lorenz_non-markovian_2005, bangert_high-resolution_2022, dai_two-dimensional_2012, jonas_two-dimensional_2003}. 
These methods have been recently successfully applied to dilute gas-phase systems\,\cite{lorenz_non-markovian_2005, dai_two-dimensional_2012, bruder_coherent_2018, roeding_coherent_2018,  bruder_delocalized_2019} and even to ultracold trapped atoms\,\cite{landmesser_two-dimensional_2023, liang_optical_2022}, which provides new perspectives to study ultracold quantum systems. 

However by nature, fs spectroscopy provides low frequency resolution which thus leaves an important asset of ultracold samples, that are the narrow spectral lines, unused. 
Hence, in order to fully exploit the potential of this novel combination of techniques, it is paramount to improve the attainable spectral resolution. 
This is possible with Fourier transform (FT) spectroscopy schemes as, for instance, integrated in CMDS and MQC experiments\,\cite{fuller_experimental_2015}.  

The above mentioned first demonstrations of nonlinear fs spectroscopy of gaseous and ultracold samples have been mostly facilitated by a special phase modulation (PM) technique, which combines rapid acousto-optic PM with efficient lock-in amplification\,\cite{tekavec_fluorescence-detected_2007}. 
This method provides both exceptional sensitivity and highly efficient action-detection\,\cite{tekavec_fluorescence-detected_2007, bruder_coherent_2018, uhl_coherent_2021} and can be readily extended to various higher-order nonlinear multi-pulse spectroscopy schemes optimized to extract specific nonlinear signals\,\cite{tekavec_fluorescence-detected_2007, bruder_delocalized_2019}. 
To attain the optimum time-frequency resolution, the PM approach incorporates FT spectroscopy principles, ranging from single (one-dimensional) interferometers to nested multidimensional interferometers. 
In this approach the spectral resolution is thus limited by the available scan range of the underlying interferometer and the acquisition speed of an interferogram. 

Previous PM experiments mostly employed a step-wise scanning protocol of the FT spectrometer units\,\cite{tekavec_fluorescence-detected_2007}. 
In this case the relative optical path length between the interferometer arms is incremented in discrete, equidistant steps. 
This approach is slow and thus limits the practically attainable resolution. 
In addition, experimental drifts are likely to occur during long acquisition times leading to spectral distortions and a reduced signal-to-noise ratio (SNR). 
This can be solved with a rapid-scanning approach recently developed for the PM technique\,\cite{agathangelou_phase-modulated_2021, autry_single-scan_2019}. 
Here, a rapid continuous scanning of the mechanical translation stage is applied combined with real-time tracking of the phase between the interferometer arms along with the interferogram. 

So far, the rapid-scanning concept has been applied to condensed phase samples where the spectral resolution is limited to the $\sim$\,THz regime by the strongly broadened response of the sample. 
In the current study, we investigate the practical resolution limit of this approach in view of high-resolution nonlinear spectroscopy in ultracold quantum systems. 
To this end, we apply rapid-scanning PM FT interferometry to ultracold atoms trapped in a magneto-optical trap (MOT), which provides a sample featuring exceptionally narrow spectral lines.   
We compare this scheme with the step-wise scanning procedure and show both the resolution and SNR advantage of the rapid-scanning approach. 
%In particular, in previous MQC and CMDS experiments of atoms trapped in a MOT, the high acquisition time in combination with limited long term stability of the complex experimental setup limited the attainable resolution to 3\,GHz\,\cite{landmesser_two-dimensional_2023}. 
With the rapid-scanning scheme we achieve a reslution of 250\,MHz, which is more then a 10-fold improvement compared to the previous experiments in a MOT\,\cite{landmesser_two-dimensional_2023}. 
%This opens the door to advanced nonlinear spectroscopy studies of ultracold quantum systems with high spectral resolution. 

\paragraph{Experimental setup}
\begin{figure}
\center
\includegraphics[scale=0.9]{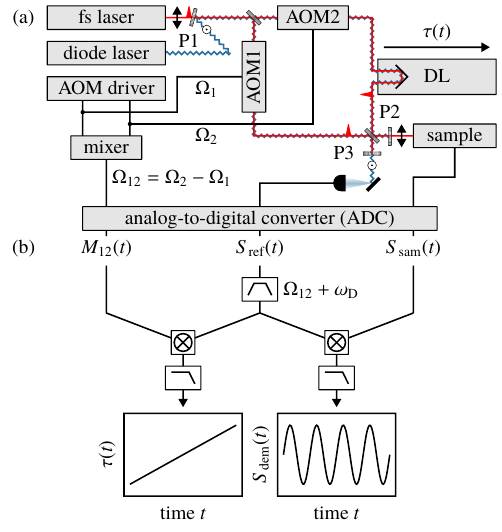}%
\caption{
\label{figure1}%
Experimental scheme: (a) optical setup, (b) signal processing. 
See main text for details.
}
\end{figure}
The PM technique has been described in detail before\,\cite{tekavec_fluorescence-detected_2007}, including the rapid-scanning method\,\cite{agathangelou_phase-modulated_2021}. 
We therefore restrict our discussion to the most essential facts. 
In most FT interferometers the optical absorption is measured. 
In contrast, the PM approach incorporates action detection. 
Here, the optical interference is induced in the sample causing a modulation of the action signal. 
Examples for action signals are the light-induced the fluorescence, photocurrent or photoionization yield of the sample\,\cite{tekavec_fluorescence-detected_2007, karki_coherent_2014, bruder_coherent_2018, uhl_coherent_2021}. 

A schematic of the experimental setup is shown in Fig.~\ref{figure1}a.
Briefly, a Mach-Zehnder interferometer equipped with a mechanical delay line (DL) is used to create a pair of collinear pulses from a fs laser system ($80\,\mathrm{MHz}$ repetition rate, $200\,\mathrm{fs}$ pulse duration).
Two acousto-optic modulators (AOM1 and AOM2) are placed in each interferometer arm,
which are driven phase-locked at frequencies $\Omega_1 = 155\,\mathrm{MHz}$ and $\Omega_2 = 155.005\,\mathrm{MHz}$, respectively, resulting in a quasi-continuous PM at frequency $\Omega_{12} = 5\,\mathrm{kHz}$. 

To achieve high spectral resolution we implement two things: a large delay range by folding the beampath traveling on the delay stage (travel range: 300\,mm) twice. 
In addition, a temperature and current-stabilized continuous-wave diode laser (frequency $\omega_\text{ref}/2 \pi \approx 384\,\mathrm{THz}$) is superimposed with the fs laser to generate an optical reference signal $S_\text{ref}$ with sufficiently long dephasing time. 
$S_\text{ref}$ is used for precise tracking of the phase of the interferometer, which in turn is used for phase-synchronous lock-in detection and for the real-time tracking of the delay $\tau$ between the two laser pulses (details below).
Wire-grid polarizers (P1-P3) are used to superimpose the diode laser with the fs laser and to separate them after passing through the interferometer before entering the sample (P2) as well as before recording the diode laser reference signal (P3).

In the step-wise scanning scheme, the delay $\tau$ between the two interferometer arms is incremented in equidistant steps and at each step the signal from the sample $S_\text{sam}$ and the optical interference of the interferometer $S_\text{ref}$ is measured, both modulated at the beat note $\Omega_{12}$. 
$S_\text{ref}$ serves as a reference for tracking the phase and phase noise of the interferometer. 
In the rapid-scanning scheme, the delay axis $\tau(t)$ is continuously scanned in real time $t$.
In order to reconstruct $\tau(t)$,
the frequency difference of the electronic waveforms driving the AOMs is generated with an analog radio-frequency mixer, yielding the signal $M_{12}(t)$. 
$M_{12}(t)$ is synchronously digitized (sampling rate $f_s = 83.3\,\mathrm{kHz}$) along with the modulated signals from the fluorescence of the sample $S_\text{sam}(t)$ and from the reference laser $S_\text{ref}(t)$. 
 
The subsequent signal processing steps are schematically shown in Fig.~\ref{figure1}b.
The delay line movement induces a velocity dependent Doppler shift $\omega_\text{D}$ in the interference signal of $S_\text{sam}(t)$ and $S_\text{ref}(t)$.
$S_\text{ref}(t)$ is filtered around the resulting oscillation at $\Omega_{12} + \omega_\text{D}$
using time-domain Butterworth filters which are preferred over Fourier filters due to computational advantages in the processing of large data arrays.
The filtered $S_\text{ref}(t)$ is used as a reference for demodulation of the signal $S_\text{sam}$ (Fig.~\ref{figure1}b right),
which is analogous to the step-wise scanning PM method\,\cite{tekavec_fluorescence-detected_2007}. 
To this end, a Hilbert transform of the filtered $S_\text{ref}(t)$ is calculated, which is then normalized and complex conjugated. 
The result is multiplied with $S_\text{sam}$ followed by a low-pass filter, which eventually yields the complex-valued demodulated data.

In case of rapid-scanning, in addition, the time-dependent delay $\tau(t)$ is reconstructed from the phase $\phi_\text{ref}(t)$ of $S_\text{ref}(t)$ obtained upon demodulation with respect to $M_{12}(t)$ (Fig.~\ref{figure1}b left). 
Using the known frequency of the reference laser $\omega_\text{ref}$, the real-time delay is computed as $\tau(t) = c_\text{ref}/c_\text{sam} \cdot \phi_\text{ref}(t) / \omega_\text{ref}$, where we assume for the ratio of the speed of light at the diode and fs laser wavelengths $c_\text{ref}/c_\text{sam} = 1$.
Importantly, with this procedure inaccuracies in the delay stage movement are effectively corrected if using a sufficiently fast data sampling:
The effective delay step size $\Delta\tau$ is given by the sampling rate $f_s$ and the scanning speed $v_\text{sc} \sim 10\,\mathrm{ps}/\mathrm{s}$ as $\Delta\tau =  v_\text{sc} / f_s \sim 100\,\mathrm{as} \ll \pi/\omega_\text{ref}$.
After demodulation, $\tau(t)$ and $S_\text{sam}$ are binned to a step size of $1\,\mathrm{fs}$. 

In both methods, $S_\text{sam}$ is zero-padded, and eventually Fourier-transformed. 
The real-part of the Fourier transform corresponds to the action-detected absorption spectrum and the imaginary part to the dispersion of the sample, respectively. 
We note, that a phase error in the interferogram leads to a mixing of both contributions and thus to a distortion of the line shape\,\cite{tekavec_fluorescence-detected_2007}. 
This can be avoided by considering just the absolute value of the Fourier spectrum, albeit at the cost of a factor of $\sqrt{2}$ lower spectral resolution. 

\begin{figure}[t]
\centering\includegraphics[scale=0.9]{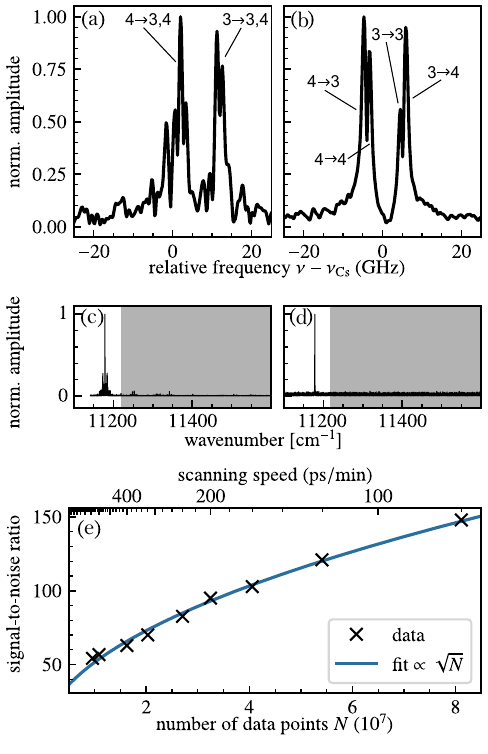}
\caption{
\label{figure2}%
FT spectra of Cs vapor recorded with the step-wise (a) and rapid-scanning (b) of the optical interferometer ($\nu_\text{Cs} = 335.116\,\mathrm{THz}$). 
Labels indicate the hyperfine transitions. 
(c,d) shows a zoom-out of the data of (a,b). 
The area dominated by white noise is marked in grey. 
(a-d) show the absolute value of the Fourier spectrum. 
(e) SNR of the rapid-scanning measurements as a function of scanning speed (top axis) and the corresponding number of data points digitized during the scan (bottom axis). 
}
\end{figure}
An important difference between the step-wise and the rapid-scanning scheme is the way how phase noise of the interferometer affects the demodulated signal. 
For interferometry in the visible spectral domain, the major noise source are usually pathlength fluctuations in the interferometer, denoted $\delta \tau$. 
These fluctuations affect the recorded time-domain interferogram $S(\tau)$ as
\begin{equation}
    S(\tau)\propto \cos \left[ \omega_\mathrm{sam} (\tau + \delta \tau) \right] \, .  
\end{equation}
Thus they scale with the resonance frequency $\omega_\mathrm{sam}$ of the sample. 
Hence, for a low-noise Fourier spectrum the fluctuations must be $\delta \tau \ll 2\pi/\omega_\mathrm{sam}$, which is difficult to achieve for spectroscopy in the VIS or UV spectral range. 
In the step-wise scanning scheme, the $\delta \tau$ fluctuations are neither explicitly measured nor actively corrected. 
However, as a result of the phase-synchronous lock-in detection, the phase difference between the interferograms $S_\text{sam}$ and $S_\text{ref}$ are actually recorded. 
The resulting interferogram after the lock-in demodulation is then
\begin{equation}
    \bar{S}(\tau)\propto \cos \left[ \Delta \omega (\tau + \delta \tau) \right] \, ,  
\end{equation}
with $\Delta \omega = \omega_\mathrm{sam} - \omega_\mathrm{ref} \ll \omega_\mathrm{sam}$. 
The phase noise scales now with $\Delta \omega \delta \tau$ and is thus reduced by several orders of magnitude, explaining the stability advantage obtained with the PM technique.  
The versatility and high efficiency of this stabilization scheme has been confirmed in various applications extending up to FT interferometry in the extreme ultraviolet spectral range\,\cite{wituschek_tracking_2020}. 

The same passive stabilization advantage applies to the rapid-scanning scheme since the same phase-synchronous detection is applied. 
However here, in addition the delay $\tau(t)$ is tracked in real-time in a separate data processing routine. 
$\tau(t)$ is then binned in the post processing to equidistant time bins which corresponds to an additional active correction of the $\delta \tau$ fluctuations. 
Consequently, the delay noise is further improved by how accurately the delay $\tau(t)$ is tracked in real time. 
The latter depends on the stability of the diode laser and can be extremely precise\,\cite{autry_single-scan_2019}.

\paragraph{Results}
We now discuss the performance differences of the step-wise and rapid-scanning scheme with the focus on experiments with high spectral resolution, where also subtle differences can be revealed. 
To this end, we performed action-detected FT spectroscopy of the D1 line ($6^{2}P_{1/2} \leftrightarrow 6^{2}S_{1/2}$ transition) in an atomic Cs vapor contained in a vapor cell at room temperature and measured the interferograms upon fluorescence detection (Fig.~\ref{figure2}). 
The electronic states involved in the Cs D1 line are the hyperfine states $F'=3,4$ and $F=3,4$ of the electronic ground and excited state, respectively. 
Accordingly, the excitation spectrum splits up into two doublets (one for $F=3 \rightarrow F'=3,4$ and one for $F=4 \rightarrow F'=3,4$). 
The spectral lines within each doublet are separated by 1.17\,GHz. 
The Doppler and the lifetime broadening of the spectral lines amount to a line width of 357\,MHz. 

To achieve the necessary resolution, we applied a scan range of 1 - 1300\,ps corresponding to an instrument response function of $0.8\,\mathrm{GHz}$. 
A challenge is to record a clean $S_\mathrm{ref}$ over the whole scan range with a resolution and frequency stability $\ll 0.8\,\mathrm{GHz}$. 
In the PM technique $S_\mathrm{ref}$ is typically obtained by spectral filtering the fs laser beams with a monochromator\,\cite{tekavec_fluorescence-detected_2007}. 
This has the advantage, that the central frequency of $S_\mathrm{ref}$ can be flexibly tuned close to the transition frequency of the sample and thus the impact of the phase noise $\Delta \omega \delta \tau$ can be minimized. 
However, the required high resolution is very challenging to achieve with an optical monochromator and thus a stabilized narrow-bandwidth diode laser is used instead to track and record $S_\mathrm{ref}$ in our experiments. 
This in turn makes tuning of the reference central frequency impractical. 
In our case the diode laser frequency is $\omega_\mathrm{ref}/2\pi=384.001$\,THz and the Cs transition is $\omega_\text{sam}/2 \pi = 335.116\,\mathrm{THz}$, thus the period of the demodulated interferogram $\bar{S}$ is $2\pi/\Delta \omega = 20.5$\,fs. 

In the step-wise scanning measurement, we chose a delay increment of $\Delta\tau = 30\,\mathrm{fs}$ and unwrap the aliased spectrum to recover the correct frequency spectrum. 
This procedure minimizes the required steps to scan the full delay range. 
Still the measurement time for the data shown in Fig.\,\ref{figure2}a,c was 13\,h. 
Using the rapid-scanning technique with a delay stage velocity set to $1.5\,\mathrm{mm/s}$, the same delay range was covered in $130\,\mathrm{s}$, being thus about 350 times faster.
In this case, aliasing is readily avoided due to the inherently small step size ($\Delta\tau < \pi / \omega_\text{ref})$,
and a binning of the time axis to $\qty{1}{\fs}$ in the post processing. 

The Fourier spectra obtained with the two methods are compared in Fig.~\ref{figure2}a-d. 
Both measurements have the same instrument response function. 
However, the hyperfine structure of the D1 line (Fig.\,\ref{figure2}a,b) is only well resolved in the rapid-scanning measurement. 
Whereas in the step-wise scanning measurement, the substructure of the hyperfine doublets is obscured by the noise. 
At the high spectral resolution of the measurement, already small irregularities in the delay axis lead to a noisy peak structure around the main resonances, which is absent in the rapid-scanning scheme where the delay axis is precisely tracked and corrected. 
We hence obtain a much better signal quality with the rapid-scanning approach despite a much shorter measurement time and thus much fewer statistics in this data set. 
This effect is only apparent in measurements with very high spectral resolution and scales with increasing $\Delta \omega$. 
For comparison, we repeated the same measurement for the Rb D2 line (not shown), where $\omega_\mathrm{ref}/2 \pi=384.001$\,THz is much closer to the optical transition of $\omega_\mathrm{sam}/2 \pi=384.2$\,THz, and thus, $\Delta \omega$ is much smaller. 
In this case the peak shapes obtained for the setp-wise and rapid-scanning measurement are basically identical. 

We note, that the main noise source apparent in Fig.~\ref{figure2}a,b is not of white-noise character. 
In a next step we investigate the influence of the latter in more detail. 
To this end, we evaluate the noise floor of both spectra reasonably far away from the resonant transitions where the noise floor is dominated by statistical white noise (Fig.\,\ref{figure2}c,d). 
Apparently, the step-wise scanning scheme produces a smaller statistical noise floor. 
For a quantitative analysis, we computed the overall SNR of the Fourier spectra by dividing the maximum amplitude of the spectrum by the RMS value of the noise floor evaluated in the gray area of Fig.\,\ref{figure2}c,d. 
In this case we obtain a much higher SNR of 500 for the step-wise scanning compared to the rapid-scanning (SNR=55). 
This is expected from the known scaling of statistical noise as $1/\sqrt{N}$  
with the total number of data points $N$ recorded for a measurment. 
Obviously, for the rapid-scanning measurement much fewer overall data points and thus much fewer statistics, are recorded. 
Fig.~\ref{figure2}e shows the SNR as a function of $N$ for the rapid-scanning scheme, which confirms the correct scaling behavior for the statistical noise. 
Here, we varied $N$ by changing the scan speed of the delay stage, which also confirms, that the scan speed has no influence on the performance of the measurement other than affecting the statistics. 
In summary, this comparison shows the performance advantage of the rapid-scanning approach in case of high-resolution measurements. 
However, it also shows, as expected, that a certain amount of statistics is essential to achieve the targeted SNR. 

\begin{figure}
\centering\includegraphics[scale=0.9]{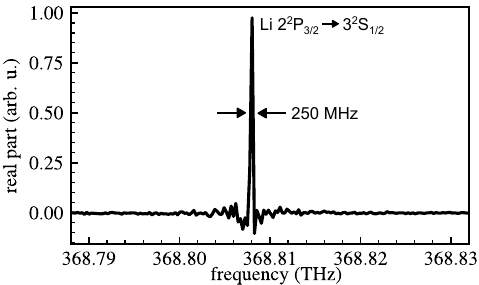}
\caption{
\label{figure3}%
FT spectrum (real part) of laser-cooled Li atoms. Labels indicate the atomic resonance and the FWHM of the spectral line.
}
\end{figure}
Having confirmed the SNR advantage of the rapid-scanning scheme in high-resolution measurements, we now focus on the practical resolution limit achievable with the rapid scanning scheme. 
To this end, we perform action-detected FT spectroscopy of laser-cooled Li atoms. 
The atoms are trapped and cooled down to $T \approx 0.8$\,mK in a magneto-optical trap (details in Ref.\,\cite{landmesser_two-dimensional_2023}). 
We study the Li $2^2P_{3/2} \rightarrow 3^2S_{1/2}$ transition ($\omega_\mathrm{sam}/ 2 \pi = 368.81$\,THz, Doppler width: 2.8\,MHz). 
Due to the small Doppler width, the sample is an ideal test system to explore the resolution limit of the method. 
The short acquisition time in the rapid-scanning approach allows us to increase the scan range to 3.9\,ns and thus reduce the instrument response function to 256\,MHz. 
Fig.~\ref{figure3} shows the obtained Fourier spectrum for fluorescence detection. 
Here, we plot the real part of the complex-valued Fourier spectrum as opposed to the absolute value plotted in Fig.\,\ref{figure2}. 
This explains why some of the side lobes around the resonance extend to negative amplitude values. 
We note, that at the high spectral resolution of the data, correct phase calibration of the data can be particularly challenging , since small calibration errors already lead to a discernible mixing of absorptive and dispersive line shapes. 

Despite the mentioned challenges in the high-resolution measurement, we obtain very good data quality, especially considering the high complexity of the experimental setup. 
The data quality is also much better than in an identical previous experiment using the step-wise scanning (Ref.\,\cite{landmesser_two-dimensional_2023}). 
The full width at half maximum (FWHM) of the measured resonance is 250\,MHz, in good agreement with the theoretical instrument response function. 
This is more than a 10-fold improvement compared to the previous experiment, where experimental drifts and the acquisition speed limited the resolution to 2.9\,GHz. 
This confirms, the feasibility of high resolution measurements in the regime well below 1\,GHz made possible by a rapid-scanning PM FT spectroscopy approach. 

In conclusion, by combining the PM technique with a rapid-scanning approach, we showed, that femtosecond FT spectroscopy of ultracold quantum systems with a resolution of 250\,MHz and very good SNR are feasible. 
The approach can be readily extended to a range of nonlinear spectroscopy schemes\,\cite{agathangelou_phase-modulated_2021, bruder_coherent_2019} ideally suited to study coherent and cooperative effects in ultracold quantum systems. 
Potentially, even higher spectral resolution could be achieved with further folding of the delay line in the interferometer or with frequency-comb technology\,\cite{lomsadze_tri-comb_2018}. 
This opens the door to advanced nonlinear spectroscopy studies of ultracold quantum systems with high spectral resolution. 
%The high spectral resolution would be beneficial e.g. in resolving individual Rydberg states in strongly interacting Rydberg ensembles while the excitation with fs laser pulses would lift the Rydberg blockade\,\cite{takei_direct_2016}, thus providing access to new interaction regimes. 
%Potentially, even higher spectral resolution could be achieved with further folding of the delay line in the interferometer or with frequency-comb technology\,\cite{lomsadze_tri-comb_2018}. 

\begin{backmatter}

\bmsection{Funding} 

\bmsection{Acknowledgments}
We thank Katrin Dulitz and Frank Stienkemeier for providing the Li-MOT setup. The following funding is acknowledged: European Research Council (ERC) Starting Grant MULTIPLEX (101078689), Deutsche Forschungsgemeinschaft (DFG) RTG 2717, Baden-Württemberg Stiftung Eliteprogram for Postdocs.

\bmsection{Disclosures}
The authors declare no conflicts of interest.

\bmsection{Data availability}
The data included in this work will be made available on the open repository Zenodo. 

\end{backmatter}

% Bibliography
\bibliography{2024RapidScanMOT}

% Full bibliography added automatically for Optics Letters submissions; the following line will simply be ignored if submitting to other journals.
% Note that this extra page will not count against page length
\bibliographyfullrefs{2024RapidScanMOT}

\end{document}